# Deep Learning for scalp High Frequency Oscillations Identification

Gaëlle Milon-Harnois
*Laboratoire Angevin en Recherche des Systèmes (LARIS)*
Angers, France
gmilonha@uco.fr

Nisrine Jrad
*Laboratoire Angevin en Recherche des Systèmes (LARIS)*
Angers, France
njrad@uco.fr

Daniel Schang
*ESEO Tech, LERIA*
Angers, France
Daniel.SCHANG@eseo.fr

Patrick Van Bogeart
*Centre Hospitalier Universitaire*
Angers, France
patrick.vanbogaert@chu-angers.fr

Pierre Chauvet
*Laboratoire Angevin en Recherche des Systèmes (LARIS)*
Angers, France
chauvet@uco.fr

*Abstract*— Since last 2 decades, High Frequency Oscillations (HFOs) are studied as a promising biomarker to localize the epileptogenic zone of patients with refractory focal epilepsy. As HFOs visual detection is time consuming and subjective, automatization of HFO detection is required. Most HFO detectors were developed on invasive electroencephalograms (iEEG) whereas scalp electroencephalograms (EEG) are used in clinical routine. In order HFO detection can benefit to more patients, scalp HFO detectors has to be developed. However, HFOs identification in scalp EEG is more challenging than in iEEG since scalp HFOs are of lower rate, lower amplitude and more likely to be corrupted by several sources of artefacts than iEEG HFOs. The main goal of this study is to explore the ability of deep learning architecture to identify scalp HFOs from the remaining EEG signal. Hence, a binary classification Convolutional Neural Network (CNN) is learned to analyze High Density Electroencephalograms (HD-EEG). EEG signals are first mapped into a 2D time-frequency image, several color definitions are then used as an input for the CNN. Experimental results show that deep learning allows simple end-to-end learning of preprocessing, feature extraction and classification modules while reaching competitive performance.

Keywords— epilepsy, High-Density Electroencephalogram (HD-EEG), High Frequency Oscillations, HFO, Convolutional Neural Network (CNN)

## I. Introduction

Epilepsy is a one of the most chronic neurological disorder concerning around 50 million patients worldwide. While current treatments are efficient to control seizures in a large majority of the patients, around 20 to 30% patients are pharmacoresistant. Some of those refractory epilepsies are due to brain lesions that can be resected to make the patient seizure-free. When epilepsy surgery is considered, an accurate localization of the brain region responsible for seizures (epileptogenic zone - EZ) is required. However, localizing the EZ is a real challenge because it does not always fit completely with the lesion. Therefore, several biomarkers of the EZ are studied since several years: the seizure onset zone (SOZ), the interictal epileptiform discharges (IEDs) and, recently, high frequency oscillations (HFOs) [1]. HFOs are brief events (between 15 to 100 ms) with regular small-amplitude oscillations within frequency range from 80 Hz to 500 Hz clearly distinguishable from background [2]. Until now, most of the studies focused on HFOs detected on invasive EEG recording using subdural or depth electrodes implanted in the brain regions selected as potential EZ. The potential source localization search is thus restricted to predefined brain regions leading to a spatial bias because it is impossible to sample all the brain regions using intracranial electrodes. In the last decade, it has been shown that HFOs can also be detected on scalp EEG [3], [4], allowing a non-invasive and affordable approach that is more applicable clinically and thus can benefit to more patients.

A recent review of HFO detection from scalp EEG [5] confirms that these events are of interest for EZ source localization, diagnosis and prognosis of epilepsy, especially in the pediatric population. In most of the studies focusing on scalp HFO detection, recording is performed using 19 electrodes placed according to the international 10–20 system [5]. Due to a poor spatial resolution, HFOs detected with this system can result in wrong EZ localization [7]. This pitfall can be solved by using High Density Electroencephalogram (HD-EEG) recording with 64 to 256 electrodes covering the whole scalp, making source localization based on HFOs more accurate [6]. So far, only few studies reported scalp HFOs using HD-EEG [6]–[8] with, for most of them, limited number of electrodes (70 to 128) and low sampling rates (500 Hz or 600 Hz) bordering HFOs to 170 Hz according to Nyquist-Shannon theorem. To tackle these issues, 256 electrodes nets and 1 kHz sampling rate were used in the current study.

Considering HFO detection, the visual one remains the gold standard, generally performed on EEG displayed on a one second window, with 10 to 20 electrodes on raw signal and / or Pass Band signal [9]. Due to low amplitude and short duration of HFOs, visual detection requires expertise, is subjective and time consuming (estimated 10 hours for 10 minutes of EEG signal on 10 electrodes) [10]. Considering the large number of electrodes and the duration of EEG records, full visual HFO detection cannot be considered. This is even worse for visual detection of HFOs from the scalp, which is much longer, error-prone, and more difficult than detection of invasively recorded HFOs. Indeed, scalp EEG signals suffer from skull EEG mitigation, artifacts and eye or muscle movements. Thus, development of efficient automatic scalp HFO detectors is required.

As the first step of detection is a good classification, this study explores the ability of deep learning architecture to identify scalp HFOs from the remaining EEG signal. In the following, we will give a brief review of existing detectors in section II. Materials and proposed methods are detailed in section III. Section IV presents experimental results. Discussion is given in section V, and section VI brings the paper to a close and set the stage to future studies.

## II. STATE OF THE ART

Since 2002, HFO detection in iEEG was widely studied in 1D domain (time or frequency domain) and more recently in 2D time-frequency (TF) domain. Most of the methods used a long pipeline to detect HFOs. This pipeline consists of artefact rejection, filtering, feature engineering (like calculation of root mean square (RMS) [9], short-term energy [11], short-term linear length [12], complex Morlet wavelet transform [13], wavelet transform of the mother wavelet [14] or Hilbert transform [15]), then features selection and eventually a classification step that rejects false detection (artefacts and IEDs). Recent studies perform feature extraction using supervised or unsupervised machine learning techniques [16]. Very few methods were proposed to automatically detect HFOs from scalp EEG [17]–[21]. Most of them are extensions of iEEG HFO detectors, are semi supervised and requires definition of threshold(s). In [21] and [19], the automated short-term energy detector was used as presented in [22]. HFO detection was then done according both RMS standard deviation threshold and peak standard deviation threshold. The team working on [17] used the invasive HFO detector presented in [23] which is based on the instantaneous Stockwell transform power spectrum. They used amplitude threshold to find HFOs. [18] developed a non-accurate semi-automated method looking for HFOs co-occurring with spikes. The method consists in computing six features derived from Hilbert transform and thresholding each of them to select potential HFOs. A visual review was then needed to reject false detections and add missing ones. The only scalp HFO detector without need for threshold definition is reported in [24] using a semi-supervised k-means algorithm followed by a mean shift algorithm to classify suspected HFOs.

To overcome all these drawbacks, we propose here an automatic classification between HFOs (from 80 Hz to 500 Hz) and EEG signal segments outside this frequency range (non HFOs) based on time-frequency maps and deep learning. Our method doesn't require any threshold definition, no distinction is performed between Ripples (frequency within 80 to 250 Hz band) and Fast Ripples (frequency ranged between 250 and 500 Hz).

## III. MATERIAL AND METHODS

### A. HD-EEG recording

Three epileptic patients followed in Department of Pediatric Neurology, Centre Hospitalier Universitaire, Angers, France, were prospectively enrolled between May and July 2020. Inclusion criteria were age under 18 years, pharmaco-resistant focal epilepsy (more than 2 treatments properly administered ineffective on seizures), occurrence of at least one seizure per day, and epileptogenic lesion visible on MRI. After informed consent signed by one parent, HD-EEG was recorded using HydroCel Geodesic Sensor Net with 256 electrodes density connected to Net Amps 400 series amplifiers (Electrical Geodesic, Inc., Eugene, OR, U.S.A.) with interconnected electrodes and the Cz electrode as a reference. Long term monitoring sensor nets were used to obtain high-quality data during up to 24 hours. The net was carefully adjusted so that Fpz, Cz, Pz and the pre-auricular points were correctly placed according to the international 10/20 system. Then each electrode was filled with Elefix paste. Electrode-skin impedances were maintained at <50 KΩ. EEG was recorded using EGI's Net Station with 1 kHz sampling rate and 0,1 Hz High Pass filter. Approximatively 18 hours of continuous HD-EEG combined with video were recorded for each patient in the awake and sleep states.

### B. HFOs Visual Detection

Thirty minutes EEG segment was selected during video confirming sleep. HFO annotation was performed visually by two experts on raw EEG signals. One second per page signals were considered on a 10-20 montage referring to Cz electrode. Every oscillatory event with minimum 3 regular oscillations clearly distinguishable from background, and frequency above 80 Hz was marked as HFO without distinction between Ripples and Fast Ripples. Selected events were then high-pass filtered and mapped in time frequency domain to confirm visual detection on raw EEG as presented in Fig. 1. Electrode, start time, end time and duration of each HFO were registered. Within each electrode, EEG signals non-belonging to HFOs were considered Non HFOs (NHFOs) zones. NHFOs zones may contain artifacts, IEDs, low frequency brain activities, background EEG, etc.

### C. Preprocessing

Preprocessing was performed with Matlab Software. For each electrode, EEG signal was normalized and high pass filtered using a 30 order Butterworth Finite Impulse Response filter with 80 Hz cutoff frequency. For each HFO, 200 ms EEG signal centered on the middle of the event was converted to time frequency maps using Short Time Fourier Transform as shown in Fig. 1. TF mapping was also applied on NHFOs; 200 ms segments randomly selected from NHFOs zones.

Images sized 875×656 pixels were obtained and then cropped to eliminate white outlines and resized to 256×256 pixels. Each image was decomposed over the 3 color channels resulting in 5 sets of images: RGB containing images with the 3 red, green and blue channels, R, G and B containing grayscale images with respectively red, green or blue channel only and HSV converting RGB images into **H**ue (the color), **S**aturation (the greyness) and Value (the brightness). An example of HFO TF maps for each color set is presented on Fig.2.

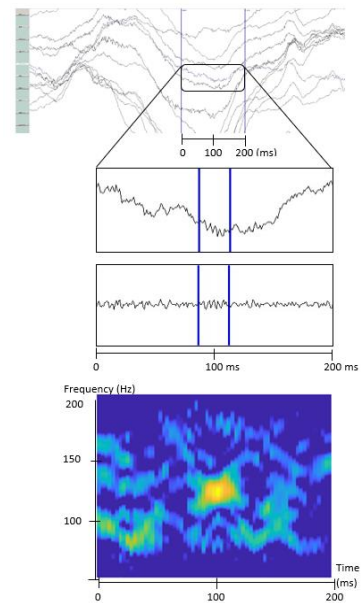

Fig. 1.    Top: 1 second raw EEG signal: blues lines point out a 200 ms time interval centered on HFO1; 2nd line: focus on selected time interval, electrode 37: blues lines marked beginning and end of HFO1; 3rd line: HFO1 on 80 Hz High Pass data; Bottom: HFO1 TF map

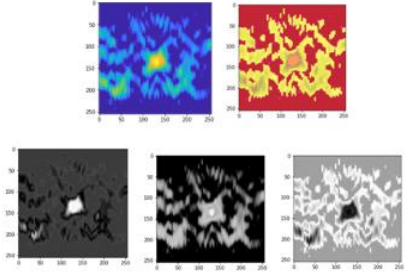

Fig. 2. Top: HFO1 TF map in RGB and HSV color map; Bottom: HFO1 TF map decomposed respectively to red (R), green (G) and blue (B) channel.

*D. CNN Architecture*

The next step consists in classifying TF maps images between HFOs and the remaining signals using a binary Convolutional Neural Network (CNN). CNN are designed to accurately recognize visual patterns directly from pixel images with minimal preprocessing [25]. CNN architectures are designed with the following 6 types of layers:

*1) The input layer* receives all the images. Each pixel of a 2D image is considered as a neuron. Mathematically, images can be represented as a tensor with 3 dimensions:

$$\dim(image) = (n_H, n_W, n_C) \quad (1)$$

with $n_H$ and $n_W$ respectively the size of height and width and $n_C$ the number of color channel(s). Our images sized 256×256 pixels and, depending the color set, contains 3 color channels for RGB and HSV and only one for R, G and B.

*2) The convolutional layers* apply filters on their input data in order to detect features of interest. A filter corresponds to a concatenation of multiple kernels which are small 2D matrix containing weights. The kernel is a squared matrix of dimension $f \times f$. The filter add a third dimension corresponding to the number of color channels. Indeed, the dimension of a filter is

$$\dim(filter) = (f, f, n_C) \quad (2)$$

Each kernel matrix is slid across the images and multiplied with the scanned part of the image. The sum of this elementwise multiplication corresponds to the cross-correlation between the image and the filter resulting in a 2D matrix. Thus, given an image I and a filter K, the cross-correlation equation for the coordinate output cell (*x*, *y*) is:

$$(I \circ K)(x,y) = \sum_{i=1}^{f}\sum_{j=1}^{f} K(i,j) \times I(x+i-1, y+j-1) \quad (3)$$

The output matrix size is:

$$\dim(output) = (n_H - f + 1, n_W - f + 1, n_C) \quad (4)$$

In our CNN, 3 convolutional layers were used with respectively 16, 32 and 64 filters of (3, 3) kernels.

*3) The maxpooling layers* extract the maximum values of the features matrices output from convolutional layer. It reduces the spatial size of the representation and thus the number of computation and avoid over fitting. In our model, a maxpooling layer of (2, 2) kernel size was used after each convolutional layer.

*4) The flatten layer* stacks the values of the features matrices output from the last maxpooling layer and provides a feature 1D matrix as input to the following layer. At this stage the features extraction part of the CNN is completed.

*5) The dense layer* is a perceptron neural network responsible for the classification part of the CNN. Each neuron is fully connected to all output neurons. After receiving an input vector, this layer applies a linear combination and then an activation function with the final aim to classify the input image. It returns as output a vector of size corresponding to the number of classes in which each component represents the probability for the image input to belong to a class. Our model uses a 500 neurons dense layer.

*6) The output layer* is a layer fully connected to the previous one. It receives the result of the CNN in as many neuron as classes. Our output layer has 2 neurons.

For convolutional, dense and output layers an activation function has to be defined to switch on or not each neuron. Several non-linear activation functions can be used. In our CNN, the Rectified Linear Unit (ReLU) function was applied on convolutional and dense layers whereas a Sigmoid function was used to activate the output layer. Mathematically, ReLu function is expressed as:

$$f(x) = \max(0, x) \quad (5)$$

Sigmoïd function corresponds to:

$$f(x) = \frac{1}{1+e^{-x}}. \quad (6)$$

Fig.3 summarizes the CNN architecture used in our study leading to a total of 28 825 086 trainable parameters. Other models were tested using a different CNN architecture (3 convolutional layers with respectively 64, 32 and 16 filters giving 7 226 382 trainable parameters) or other activation functions (Leaky ReLU for convolutional layers and Sofmax for output layer). Those models results were not reported because they were lower than the ones presented.

*E. Training and testing CNN*

TF maps were randomly resampled into 3 different datasets: 64% for training, 16% for validation and 20% for testing. Training of the CNN model was performed on the training set images. Each TF map belonging to this dataset was processed by batches of 20 images in 10 epochs unless early stopping if loss accuracy does not improve after 4 epochs. Number of steps by epoch corresponds to the number of samples divided by the number of batches. In case of early stopping, model weights are restored from the end of the best epoch. Validation dataset was used to compute the accuracy and the loss of the model at each epoch. Performance metrics were evaluated on test dataset from the best epoch of the classification model. In order to check the robustness of our model, 12 runs on as much different random selections of images in the training, validation and test datasets were performed. Machine learning was executed using mostly tensorflow and keras packages in Google Colab, a Google cloud service, based on Jupyter Notebook and intended for training and research in machine learning.

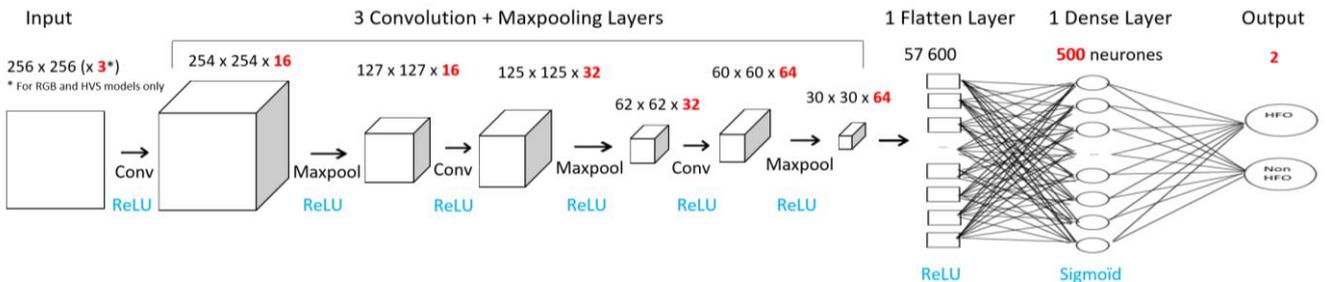

Fig. 3: Detail of the Convolutional Neural Network model used to classify HFO / NHFO

### F. Performance metrics

In order to ensure model quality, the following metrics were calculated:

- Recall or sensitivity (Se) measuring true positive rate :

$$Se = TP/(TP + FN) \quad (7)$$

- Precision or Positive Predictive Value (PPV):

$$PPV = TP/(TP + FP) \quad (8)$$

- Specificity (Spe) corresponding to true negative rate:

$$Spe = TN/(TN + FP) \quad (9)$$

- F1-score measuring model ability to properly predict HFOs in term of both precision and recall. It corresponds to harmonic means between precision and recall:

$$F1score = TP/[TP + 0.5 \times (FN + FP)] \quad (10)$$

Where true positive (TP) refers to the visually marked HFOs recognized by the CNN model, false positive (FP) corresponds to TF maps classified HFOs but not visually marked, false negative (FN) are visually marked HFOs missed by the classifier and true negative (TN) refers to NHFOs properly classified by the model.

## IV. EXPERIMENTAL RESULTS

Our dataset of 5182 images equally spread between HFOs and NHFOs was randomly separated in 1036 images in Test set and 4146 images for model learning subdivided in 3318 images in the training set and 828 in the validation set equally distributed between the 2 classes as shown in Fig.4.
Means and standard errors of metrics resulting from the 12 runs was summarized in table I by set of images. Model trained on HSV color space and blue channel returns lower results than the other color sets. Whereas models running on 3 dimensional RGB images and on greyscale images on red or green channel returns very good precision, recall and F1 score ranging from 81% to 86% with low variability (from 1% to 5%). Specificity is higher and more robust for RGB images (86%±2%) than for the other color sets.

TABLE I. METRICS SUMMARY (MEAN ± STANDARD ERROR)

|  | RGB | R | G | B | HSV |
|---|---|---|---|---|---|
| Precision | 86%±2% | 83%±3% | 81%±4% | 81%±2% | 79%±4% |
| Recall | 85%±3% | 86%±3% | 85%±5% | 81%±3% | 85%±5% |
| Specificity | 86%±2% | 82%±4% | 80%±6% | 81%±3% | 77%±6% |
| F1-score | 85%±1% | 84%±1% | 83%±2% | 81%±1% | 82%±2% |

## V. DISCUSSION

We show here that CNN with 3 convolutional layers can be considered as an efficient classifier discriminating HFOs and NHFOs TF maps presented as RGB or red grayscale images. Results on green channel are mitigated. While results obtained with RGB and R color definitions seem very promising, conversion of TF images to HSV color space and blue channel seems less interesting.

On several HFO classification studies performed on invasive EEG, recalls are very close to our results: [26] reported 87.4% sensitivity with a 2 steps traditional machine learning process. Similarly, authors of [27], used a CNN preceded by Short-Time Energy to extract features. They reported 88.16% recall for Ripples and 93.37% for Fast Ripples. Since EEG signals we studied were recorded using 1 kHz sampling rate, most of HFOs we detected are more likely to be Ripples or not very high frequencies Fast Ripples. Thus, when comparing to iEEG Ripples recall reported in [27], we obtained similar performance using all color sets except blue. Precision (88.67%) and F1 score (89.95%) reported in this latter study are also similar to our RGB results. Another iEEG HFO detection study using CNN [28] obtained 77.04% recall for ripples and 83.23% recall for fast ripples, which are lower than our results on all color sets except blue one.

Very few studies were done on scalp EEG. We will compare our results to studies that reported performance metrics for HFO automatic detection. Only 68.2% sensitivity was reported on [24] whereas 55% ±15 % sensitivity was calculated on 3 patients in [21]. It is worthy to note that our results are much more robust to variability than those of [21].

Results obtained in our study are hence comparable to the ones resulting from iEEG studies and better than existing scalp HFO automatic detection. Moreover, our model is simple using a single step for feature extraction and classification and with signal preprocessing limited to normalization, High Pass filtering and Time Frequency maps generation. Considering that our HFO classification was performed on scalp EEG with a more tedious detection due to lower signal intensity and more artifacted signals, results obtained are very promising. Moreover, results on RGB (3 channels) and R images are good, which means that computation time can be significantly reduced using a one color channel. Since our NHFO events were randomly selected, we are confident that they are representative of all kind of activities recorded in scalp EEG and that results will remain competitive when extending our detection on the whole electrodes.

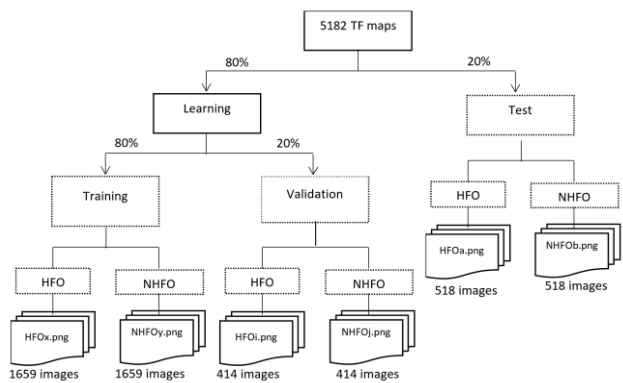

Fig. 4 Resampling images into training, validation and test datasets. Dashed boxes correspond to directories containing TF maps resized images

## VI. Conclusion

Our main contribution is to prove that CNN model for feature extraction and classification gives interesting metrics on scalp EEG HFO detection. If working on HSV colored images and blue channel are not conclusive, metrics resulting from all others color sets are comparable to results obtained for iEEG HFO classification. Further analyses and models need to be tested using red and green channels in order to obtain an even better detection. Other deep learning models will be explored on 1D EEG signal in order to avoid time frequency mapping. Moreover, once we get an efficient HFO automatic detector, we will compare EZ delineated by HFOs with the ones defined with IEDs and seizures.